\documentclass{article}%
\usepackage{breckenridge}
\input psfig.sty
\frompage{000} \topage{000}
\def\rightmark{Nuclear Astrophysics in Rare Isotope Facilities}
\def\leftmark{C.A. Bertulani}
\title{{\bf Nuclear Astrophysics in Rare Isotope Facilities\\}}
\authors{
{C.A. Bertulani}\\[2.812mm]
{\normalsize \hspace*{-8pt} NSCL and Department of Physics and
Astronomy,
Michigan State University, East Lansing, MI 48824\\
[0.2ex]  }}  \abstract{I discuss a few of the recent developments
in nuclear reactions at very low energies with emphasis on the
role of radioactive beam facilities.} 
\begin{document}
\maketitle

\setcounter{page}{1}

\section{Introduction}

\label{intro}

Present studies in nuclear astrophysics are focused on the
opposite ends of the energy scale for nuclear reactions: (a) the
very high and (b) the very low relative energies between the
reacting nuclei. Projectiles with high bombarding energies produce
nuclear matter at high densities and temperatures. This is the
main goal at the RHIC accelerator at the Brookhaven National
Laboratory and also the main subject discussed in this Workshop.\
One expects that matter produced in central nuclear collisions at
RHIC for $\sim10^{4}$ GeV/nucleon of relative energy, and at the
planned Large Hadron Collider at CERN, will undergo a phase
transition and produce a quark-gluon plasma. One can thus
reproduce conditions existent in the first seconds of the universe
and also in the core of neutron stars. At the other end of the
energy scale are the low energy reactions of importance for
stellar evolution. A chain of nuclear reactions starting at
$\sim10-100$ keV leads to complicated phenomena like supernovae
explosions or  the energy production in the stars.

Nuclear astrophysics requires the knowledge of the reaction rate
$R_{ij}$ between the nuclei $i$ and $j$. It is given by
$R_{ij}=n_{i}n_{j}<\sigma v>/(1+\delta_{ij})$, where $\sigma$ is
the cross section, $v$ is the relative velocity between the
reaction partners, $n_{i}$ is the number density of the nuclide
$i$, and $<>$ stands for energy average.

In our Sun the reaction $^{7}Be\left(  p,\gamma\right)  ^{8}B$
plays a major role for the production of high energy neutrinos
originated from the $\beta$-decay of $^{8}B$. These neutrinos come
directly from center of the Sun and are an ideal probe of the
Sun's structure. Long ago, Barker \cite{Bar80} has emphasized that
an analysis of the existing experimental data yields an S-factor
for this reaction at low energies which is uncertain by as much as
30\%. This situation has changed recently, mainly due to the use
of radioactive beam facilities.  The reaction $^{12}C\left(
\alpha,\gamma\right)  ^{16}O$ is extremely relevant for the fate
of massive stars. It determines if the remnant of a supernova
explosion becomes a black-hole or a neutron star. It is argued
that the cross section for this reaction should be known to better
than 20\%, for a good modelling of the stars \cite{Woo85}. This
goal has not yet been achieved.

Both the $^{7}Be\left( p,\gamma\right) ^{8}B$ and the
$^{12}C\left( \alpha,\gamma\right) ^{16}O$ reactions\ cannot be
measured at the energies occurring inside the stars (approximately
20 keV and 300 keV, respectively). Direct experimental
measurements at low energies are often plagued with low-statistics
and large error bars. Extrapolation procedures are often needed to
obtain cross sections in the energy region of astrophysical
relevance. While non-resonant cross sections can be rather well
extrapolated to the low-energy region, the presence of continuum,
or subthreshold resonances, complicates these extrapolations.
Numerous radiative capture reactions pose the same experimental
problem.

Approximately half of all stable nuclei observed in nature in the
heavy element region about $A>60$ is produced in the r--process.
This r--process occurs in environments with large neutron
densities which lead to$\;\tau _{\mathrm{n}}\ll\tau_{\beta}$. The
most neutron--rich isotopes along the r--process path have
lifetimes of less than one second; typically 10$^{-2}$ to
10$^{-1}$\thinspace s. Cross sections for most of the nuclei
involved are hard to measure experimentally. Sometimes,
theoretical calculations of the capture cross sections as well as
the beta--decay half--lives are the only source of the nuclear
physics input for r--process calculations \cite{Sch01}. For nuclei
with about $Z>80$ beta--delayed fission and neutron--induced
fission might also become important.

\section{The Electron Screening Problem}

Besides the Coulomb barrier, nucleosynthesis in stars is
complicated by the presence of electrons. They screen the nuclear
charges, therefore increasing the fusion probability by reducing
the Coulomb repulsion. Evidently, the fusion cross sections
measured in the laboratory have to be corrected by the electron
screening when used as inputs of a stellar model. This is a purely
theoretical problem as one can not reproduce the interior of
stars in the laboratory. Applying the Debye-H\"{u}ckel, or
Salpeter's, approach \cite{Sal54}, one
finds that the plasma enhances reaction rates, e.g., $^{3}He(^{3}%
He,\ 2p)^{4}He$ and $^{7}Be(p,\ \gamma)^{8}B$, by as much as 20\%.
This does not account for the dynamic effect due to the motion of
the electrons (see, e.g., \cite{CSK88,Brown97}).

A simpler screening mechanism occurs in laboratory experiments due
to the bound atomic electrons in the nuclear targets.  This case
has been studied in great details experimentally, as one can
control different charge states of the projectile+target system in
the laboratory \cite{Ass87,Eng88,Blu90,Rol95,Rol01}. The
experimental findings disagree systematically by a factor of two
with theory. This is surprising as the theory for atomic screening
in the laboratory relies on our basic knowledge of atomic physics.
At very low energies one can use the simple adiabatic model in
which the atomic electrons rapidly adjust their orbits to the
relative motion between the nuclei prior to the fusion process.
Energy conservation requires that the larger electronic binding
(due to a larger charge of the combined system) leads to an
increase of the relative motion between the nuclei, thus
increasing the fusion cross section. As a matter of fact, this
enhancement has been observed experimentally. The measured values
are however not compatible with the adiabatic estimate
\cite{Ass87,Eng88,Blu90,Rol95,Rol01}. Dynamical calculations have
been performed, but they obviously cannot explain the discrepancy
as they include atomic excitations and ionizations which reduce
the energy available for fusion. Other small effects, like vacuum
polarization, atomic and nuclear polarizabilities, relativistic
effects, etc., have also been considered \cite{BBH97}. But  the
discrepancy between experiment and theory remains
\cite{BBH97,Rol01}.

\centerline{\psfig{figure=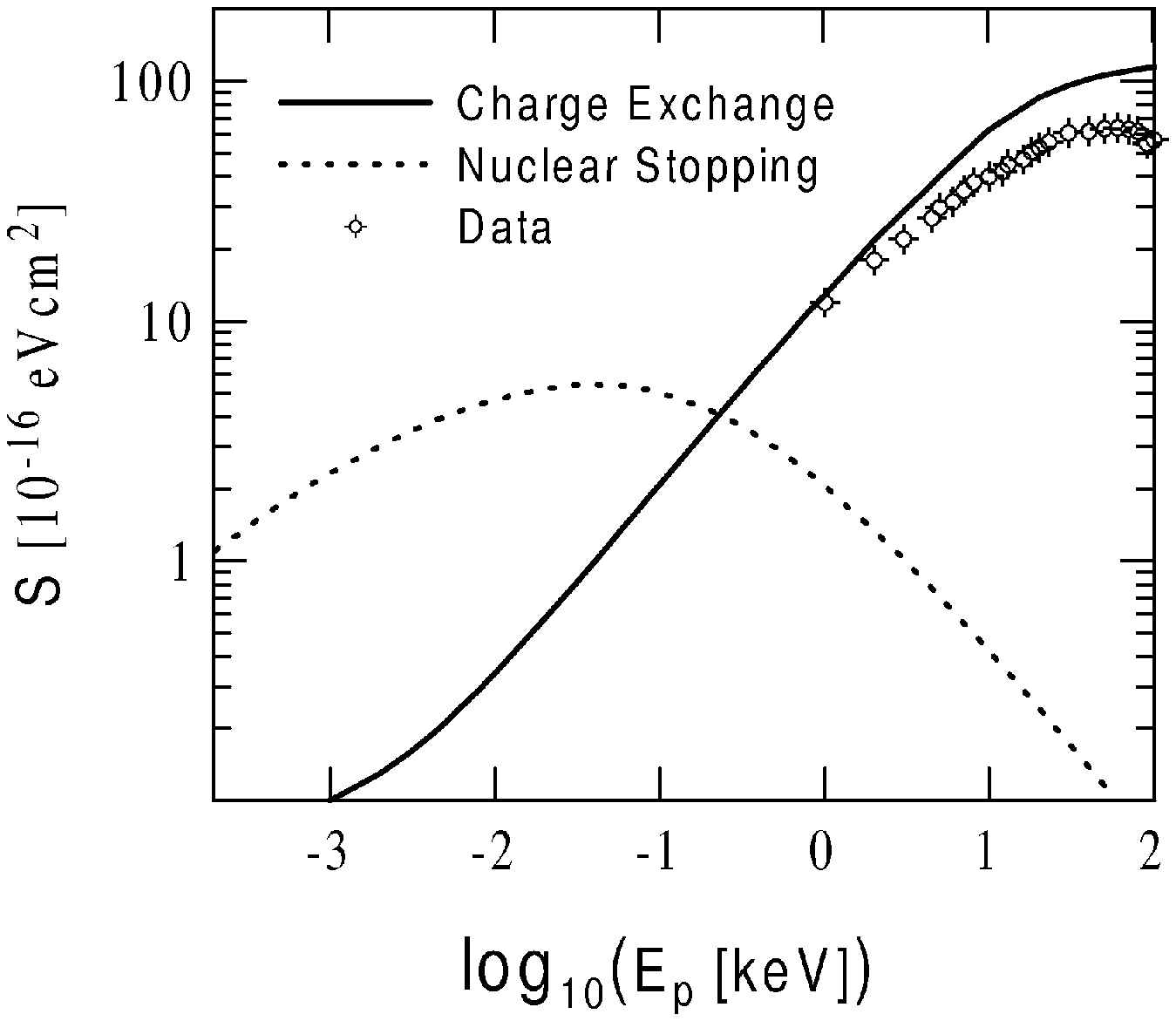,height=2in },\ \
\psfig{figure=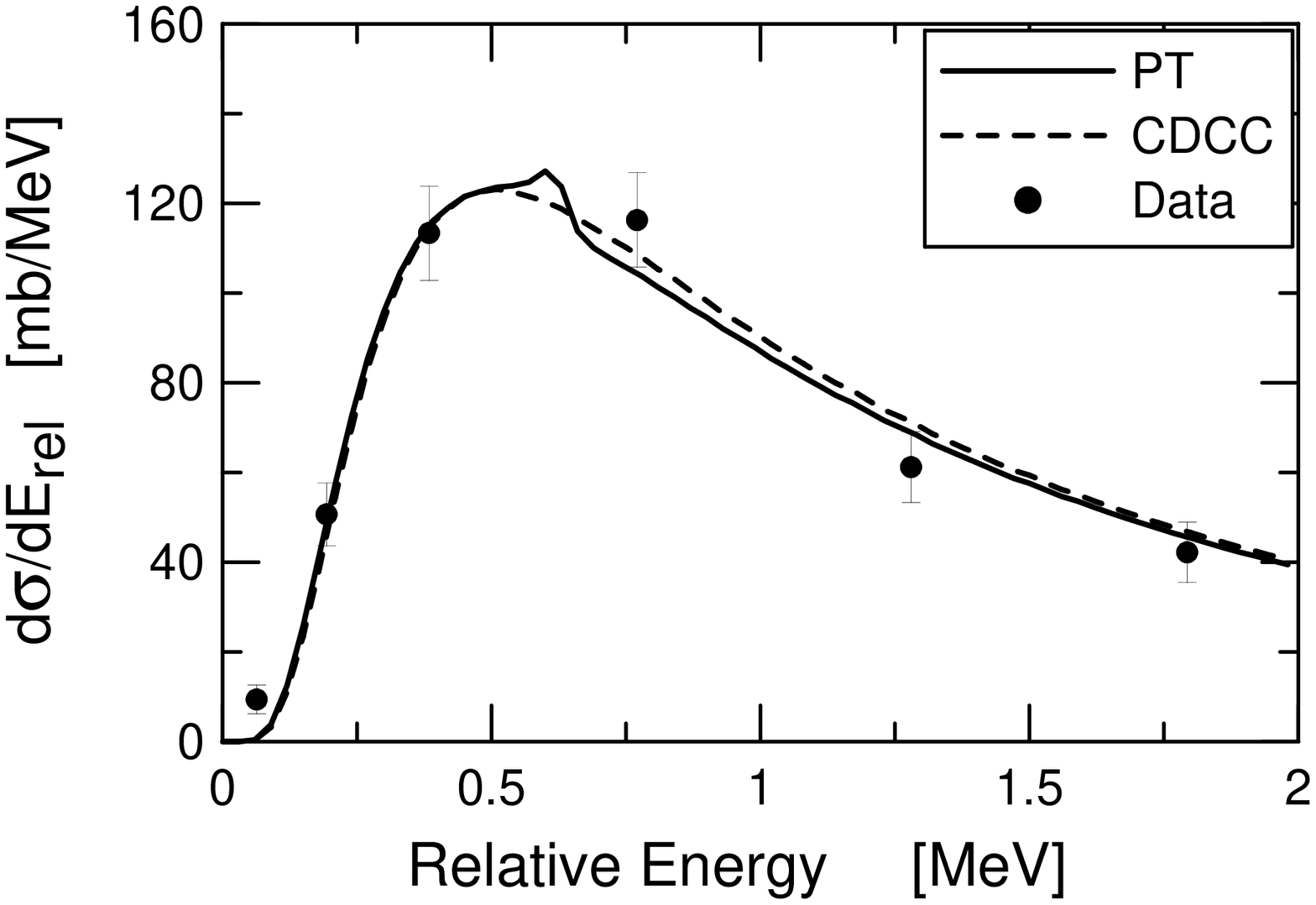,height=1.8in }}

{\small \underline{Fig. 1:} The stopping cross section of protons
on H-targets. The dotted line gives the energy transfer by means
of nuclear stopping, while the solid line is the result for the
charge-exchange stopping mechanism \cite{BD00}. The data points
are from the tabulation of Andersen and Ziegler \cite{AZ77}.

\underline{Fig. 2:}
Energy dependence of the Coulomb breakup cross section for $^{8}\mathrm{B}%
+\mathrm{Pb}\longrightarrow\mathrm{p}+^{7}\mathrm{Be}+\mathrm{Pb}$
at 84 MeV/nucleon. First-order perturbation calculations (PT) are
shown by the solid curve. The dashed curve is the result of a CDCC
calculation including the coupling between the ground state and
the low-lying states with the giant dipole and quadrupole
resonances \cite{Ber02}. The data points are from ref.
\cite{Da01}.}
\   \\

A possible solution of the laboratory screening problem was
proposed in refs. \cite{LSBR96,BFMH96}. Experimentalists often use
the extrapolation of the Andersen-Ziegler tables \cite{AZ77} to
obtain the average value of the projectile energy due to stopping
in the target material. The stopping is due to ionization,
electron-exchange, and other atomic mechanisms. However, the
extrapolation is challenged by theoretical calculations which
predict a lower stopping. Smaller stopping was indeed verified
experimentally  \cite{Rol01}. At very low energies, it is thought
that the stopping mechanism is mainly due to electron exchange
between projectile and target. This has been studied in ref.
\cite{BD00} in the simplest situation; proton+hydrogen collisions.
Two-center electronic orbitals were used as input of a
coupled-channels calculation. The final occupation amplitudes were
projected onto bound-states in the target and in the projectile.
The calculated stopping power was added to the nuclear stopping
power mechanism, i.e. to the energy loss by the Coulomb repulsion
between the nuclei. The obtained stopping power is proportional to
$v^{\alpha }$, where $v$ is the projectile velocity and
$\alpha=1.35$. The extrapolations from the Andersen-Ziegler table
predict a larger value of $\alpha$. Although this result seems to
indicate the stopping mechanism as a possible reason for the
laboratory screening problem, the theoretical calculations tend to
disagree on the power of v at low energy collisions. For example,
ref. \cite{GS91}  found $ S\sim v_{p}^{3.34}$ for protons in the
energy range of 4 keV incident on helium targets. This is an even
larger deviation from the extrapolations of the Andersen-Ziegler
tables.

We are faced here with a notorious case of obscurity in nuclear
astrophysics. The disturbing conclusion is that as long as we
cannot understand the magnitude of electron screening in stars or
in the atomic electrons in the laboratory, it will be even more
difficult to understand color screening in a quark-gluon plasma,
an important tool in relativistic heavy ion physics (e.g., the
$J/\Psi$ suppression mechanism).

\section{Radioactive Beam Facilities and Indirect Methods}

Transfer reactions are a well established tool to obtain spin,
parities, energy, and spectroscopic factors of states in a nuclear
system. Experimentally, $(d,\ p)$ reactions are mostly used due to
the simplicity of the deuteron. Variations of this method have
been proposed by several authors. For example, the {\it Trojan
Horse} Method was proposed in ref. \cite{Bau86} as a way to
overcome the Coulomb barrier.  If the Fermi momentum of the
particle $x$ inside $a=(b+x)$ compensates for the initial
projectile velocity $v_{a}$, the low energy reaction $A+x=B+c$ is
induced at very low (even vanishing) relative energy between $A$
and $x$. Successful applications of this method has been reported
recently \cite{Spit00}. Also recently the {\it knockout reactions}
have been demonstrated to be a useful tool to deduce spectroscopic
factors in many reactions of relevance for nuclear astrophysics
\cite{HJ03}.

At low energies the amplitude for the radiative capture cross
section is dominated by contributions from large relative
distances of the participating nuclei. Thus, what matters for the
calculation of the direct capture matrix elements are the {\it
asymptotic normalization coefficients} (ANC). This coefficient is
the product of the spectroscopic factor and a normalization
constant which depends on the details of the wave function in the
interior part of the potential. The normalization coefficients can
be found from peripheral transfer reactions whose amplitudes
contain the same overlap function as the amplitude of the
corresponding astrophysical radiative capture cross section. This
idea was proposed in ref. \cite{Muk90} and many successful
applications of the method have been obtained \cite{Cag}.

{\it Charge exchange} induced in $(p,n)$ reactions are often used
to obtain values of Gamow-Teller matrix elements which cannot be
extracted from beta-decay experiments. This approach relies on the
similarity in spin-isospin space of charge-exchange reactions and
$\beta$-decay operators. As a result of this similarity, the cross
section $\sigma(p,\ n)$ at small momentum transfer $q$ is closely
proportional to $B(GT)$ for strong transitions \cite{Tad87}. As
shown in ref. \cite{Aus94}, for important GT transitions whose
strength are a small fraction of the sum rule the direct
relationship between $\sigma(p, \ n)$ and $B(GT)$ values fails to
exist. Similar discrepancies have been observed \cite{Wat85} for
reactions on some odd-A nuclei including $^{13}C$, $^{15}N$,
$^{35}Cl$, and $^{39}K$ and for charge-exchange induced by heavy
ions \cite{Ber96,St96}.

The (differential, or angle integrated) {\it Coulomb breakup}
cross section for $a+A \longrightarrow b+x+A$ can be written as
$\sigma_{C}^{\pi\lambda} (\omega) = F^{\pi\lambda}(\omega) \ .\
\sigma_{\gamma}^{\pi\lambda} (\omega) $, where $\omega$ is the
energy
transferred from the relative motion to the breakup, and $\sigma_{\gamma}%
^{\pi\lambda} (\omega)$ is the photo nuclear cross section for the
multipolarity ${\pi\lambda}$ and photon energy $\omega$. The
function $F^{\pi\lambda}$ depends on $\omega$, the relative motion
energy, and nuclear charges and radii. They can be easily
calculated \cite{Ber88} for each multipolarity ${\pi\lambda}$.
Time reversal allows one to deduce the radiative capture cross
section $b+x \longrightarrow a+\gamma$ from
$\sigma_{\gamma}^{\pi\lambda} (\omega)$.  This method was proposed
in ref. \cite{BBR86}. It has been tested successfully in a number
of reactions of interest for astrophysics (\cite{Bau94} and
references therein).  The most celebrated case is the reaction
$^{7}Be(p, \gamma)^{8}B$. It has been studied in numerous
experiments in the last decade. For a recent compilation of the
results obtained with the method, see the contribution of Moshe
Gai to this workshop, and also ref. \cite{DT03}. They have
obtained an $S_{17}(0)$ value of 19.0 eV.b which is compatible
with the value commonly used in solar model calculations
\cite{Bah89}. To achieve the goal of applying this method to many
other radiative capture reactions (for a list, see, e.g.
\cite{Bau94}), detailed studies of dynamic contributions to the
breakup have to be performed, as shown in refs. \cite{BBK92,BB93}.
The role of higher multipolarities (e.g., E2 contributions
\cite{Ber94,GB95,EB96} in the reaction $^{7}Be(p, \gamma)^{8}B$)
and the coupling to high-lying states \cite{Ber02} has also to be
investigated carefully. In the later case, a recent work has shown
that the influence of giant resonance states is small (see figure
2). Studies of the role of the nuclear interaction in the breakup
process is also essential to determine if the Coulomb dissociation
method is useful for a given system \cite{BD03}.

In summary, radioactive beam facilities have opened a new paved
way to disclosure many unknown features of reactions in stars and
elsewhere in the universe.

\ \

I acknowledge discussions with B. Davids, M. Gai, H. Schatz, K.
Suemmerer and S. Typel. Work Supported by U.S. National Science
Foundation under Grants No. PHY-007091 and PHY-00-70818.


\begin{thebibliography}{99}                                                                                               %


\bibitem {Bar80}F.C. Barker, Aust. J. Phys. 33 (1980) 177; Phys. Rev. C37
(1988) 2930.

\bibitem {Woo85}S.E. Woosley, Proceedings of the Accelerated Radioactive Beam
Workshop, eds. Buchmann and J.M. D'Auria (TRIUMF, Canada, 1985).

\bibitem {Sch01} H. Schatz,et al., Phys. Rev. Lett. 86  (2001) 3471.

\bibitem {Sal54}E.E. Salpeter, Aust. J. Phys. 7 (1954) 373.

\bibitem {CSK88} C. Carrero,  A. Sch\"afer and S. E. Koonin,  Astrophys. J. 331 (1988)
565.

\bibitem{Brown97}  Lowell S. Brown, R. F. Sawyer, Rev. Mod. Phys.  69 (1997)
411.

\bibitem {Ass87}H.J. Assenbaum, K. Langanke and C. Rolfs, Z. Phys. A327 (1987)
461.

\bibitem {Eng88}S. Engstler \textit{et al.}, Phys. Lett. B202 (1988)
179.

\bibitem {Blu90}G. Bl\"{u}gge and K. Langanke, Phys. Rev. C41 (1990) 1191; K.
Langanke and D. Lukas, Ann. Phys. 1 (1992) 332.

\bibitem {Rol95}C. Rolfs and E. Somorjai, Nucl. Inst. Meth. B99 (1995)
297.

\bibitem{Rol01}C. Rolfs, Prog. Part. Nucl. Phys. 46 (2001) 23.

\bibitem{BBH97} A.B. Balantekin, C.A. Bertulani, M.S. Hussein, Nucl. Phys.
A627 (1997) 324.

\bibitem {Da01}B. Davids, {\it et al.}, Phys. Rev. C63 (2001) 065806.

\bibitem{LSBR96}  K. Langanke, T.D. Shoppa, C.A. Barnes and C. Rolfs, Phys.
Lett. {\bf B369}, 211 (1996).

\bibitem{BFMH96}  J.M. Bang, L.S. Ferreira, E. Maglione, and J.M. Hansteen,
Phys. Rev. {\bf C53}, R18 (1996).

\bibitem{AZ77} H. Andersen and J.F. Ziegler, {\it The Stopping and
Ranges of Ions in Matter}, Pergamon, NY, 1977.

\bibitem{BD00} C.A. Bertulani and D.T. de Paula, Phys. Rev. C62 (2000) 045802.


\bibitem{GS91}  R. Golser and D. Semrad, Phys. Rev. Lett. {\bf 14}, 1831
(1991)


\bibitem{Spit00}C. Spitaleri, {\it et al.}, Eur. Phys. J, A7 (2000)
181.

\bibitem{HJ03} P.G. Hansen and J.A. Tostevin, Ann. Rev. Nucl.
Part. Sci., to be published.

\bibitem {Bau86}G. Baur, Phys. Let. B178 (1986) 135.

\bibitem {Muk90}A.M. Mukhamedzhanov and N.K. Timofeyuk, JETP Lett. 51 (1990)
282.

\bibitem{Cag}  C.A. Gagliardi, {\it et al.}, Phys. Rev. C59 (1999)
1149.

\bibitem {Tad87}T.N. Tadeucci \textit{et al.}, Nucl. Phys. A469 (1987)
125.

\bibitem {Aus94}S.M. Austin, N. Anantaraman and W.G. Love, Phys. Rev. Lett. 73
(1994) 30.

\bibitem {Wat85}J.W. Watson et al., Phys. Rev. Lett. 55 (1985)
1369.

\bibitem {Ber96} C.A. Bertulani, Nucl. Phys. A554 (1993) 493;
C. A. Bertulani and P. Lotti, Phys. Lett. B402 (1997) 237;  C.A.
Bertulani and D. Dolci, Nucl. Phys. A674 (2000) 527.

\bibitem{St96} M. Steiner, {\it et al.},  Phys. Rev. Lett. 76 (1996)
26.

\bibitem {Ber88}C. Bertulani and G. Baur, Phys. Reports 163 (1988) 299

\bibitem {BBR86}G. Baur, C. Bertulani and H. Rebel, Nucl. Phys. A459 (1986)
188.

\bibitem {Bau94}G. Baur and H. Rebel, J. Phys. G20 (1994) 1

\bibitem{DT03} B. Davids, S. Typel, Los Alamos preprint server,
nucl-th/0304054.

\bibitem {Bah89}J.N. Bahcall,``Neutrino Astrophysics'', Cambridge University
Press, 1989

\bibitem{BBK92} G. Baur, C.A. Bertulani and D.M. Kalassa, Nucl. Phys. A550
(1992) 527.

\bibitem{BB93} G.F. Bertsch and C.A. Bertulani, Nucl. Phys. A556 (1993)
136; Phys. Rev. C49 (1994) 2834;  H. Esbensen, G.F. Bertsch and
C.A. Bertulani, Nucl. Phys. A581 (1995) 107.

\bibitem{Ber94}C.A. Bertulani, Phys. Rev. C49 (1994) 2688; Nucl. Phys. A587 (1995) 318;
Z. Phys. A356 (1996) 293.

\bibitem{GB95} M. Gai and C.A. Bertulani, Phys. Rev. C52 (1995) 1706;
Nucl. Phys. A636 (1998)  227

\bibitem{EB96} H. Esbensen and G. Bertsch, Phys. Rev. Lett. A359 (1995) 13;
Nucl. Phys. A600 (1996) 37.

\bibitem{Ber02} C.A. Bertulani, Phys. Lett. B3  (2002) 205.

\bibitem{BD03}C.A. Bertulani and P. Danielewicz, Nucl. Phys. A717 (2003) 199.

\end{thebibliography}
\end{document}